# The concept of particle in Quantum Field Theory

Eliano Pessa*[*]

(*Interdepartmental Center for Research in Cognitive Science, University of Pavia, Pavia, Italy*)

Despite its name, Quantum Field Theory (QFT) has been built to describe interactions between localizable particles. For this reason the actual formalism of QFT is partly based on a suitable generalization of the one already used for systems of point particles. This circumstance gives rise to a number of conceptual problems, stemming essentially from the fact that the existence within QFT of non-equivalent representations implies the existence of field theories allowing, within the same theory, different, inequivalent, descriptions of particles. This led some authors to claim that in QFT the concept itself of particle should be abandoned. In this paper we will shortly discuss the validity of this claim, as well as the possibilities, so far existing, of building alternative versions of QFT, not designed in advance to allow some kind of particle representation. We will also spend some words about the generalizations of the concept itself of particle which could grant for a better cohabitation of particles and fields within a wider formulation of QFT. The latter is indispensable if we want to extend the range of application of QFT from particle physics or simple condensed matter physics to other domains of scientific research.

## 1. Introduction

As well known, the actual formulation of QFT stems from the theoretical developments of classical physics which arose as a consequence of the controversies characterizing the second part of the nineteenth and the beginning of the twentieth century. In particular we refer to the debate between holders of particle theories and holders of field theories. Roughly speaking (see, for instance, Redhead, 1982) a particle theory can be defined as a theory which attributes to suitable individual entities (the particles) a number of properties. On the contrary a field theory associates certain properties with every space-time point. Despite the popularity of field theories, strongly increased after the discovery of electromagnetic waves, even particle theories, initially supported by Newton himself, gained a wide consensus, owing to experiments proving the validity of atomic view of matter. As it easy to understand, particle theories entail, with respect to field theories, a further problem: how to define the individual entities (that is, the particles)? Namely, as the properties of the latter (for instance, their location) can vary with time and across the space, we need a further theory specifying how to recognize in an invariant way each individual entity despite the fact that some or all of its properties have undergone a change. Such a theory, which we could call a *model of the particle*, should also tells us why a given individual entity is associated with some properties and not with others. Obviously, this explanation of the origin of individual properties could not be based, to avoid any *regressio ad infinitum*, on concepts making appeal to other kinds of individuals or to other particle theories. These requirements show that the building of a model of the particle is a very difficult task and this accounts for the small number of attempts made to reach this goal (see in this regard the classical books Whittaker, 1951, 1953, as well as Doran, 1975; a more modern reference is given by Jiménez and Campos, 1999).

The above remarks help to understand why the only model of the particle so far practically adopted by all particle theories (neglecting here string theories) is the one of *point particles*. The adoption of this model is at the very origin of a number of infinities and divergences, still plaguing physics and, in particular, QFT, all stemming from the fact that a point has zero geometrical extension. From the historical point of view the introduction of a view based on point particles owes very much to the work of H.A.Lorentz, who reformulated the original Maxwell's theory of electromagnetism in terms of pointlike sources transmitting their influence through a non-mechanical aether (for more details about the role played by Lorentz see McCormmach, 1970). It is very difficult to underestimate the importance of Lorentz's contribution: both Quantum Mechanics (QM) and QFT have been shaped from the beginning as theories about particles or quanta. Of course, within a quantum-theoretical framework the wave-particle duality helped very much in associating quanta with harmonic oscillators (and therefore most particles with suitable wavepackets). This contributed to hide, at least momentarily, the conceptual difficulties stemming from the fact that the mathematical structure of QM does nothing but generalize the one of mechanics of point particles. In any case, the general consensus about the possibility of characterizing each individual entity (each particle) by resorting only to the (supposedly) invariant properties describing its "charges" (like its inertial mass, electric charge, spin, and so on) exempted from finding a deeper explanation of the origin of charges themselves. And this attitude stopped any further investigation about the possible relationships between these charges and other particle properties of spatiotemporal nature. On the other hand, in accordance with Noether's theorem, the Hamiltonians describing systems of interacting particles are, already from the starting, chosen in such a way as to be gauge-invariant.

---

[*] e-mail: eliano.pessa@unipv.it

This quiet world, in which QM and QFT were developing by dealing with particles as if they were really "elementary" objects, broke down at the end of the Forties when people found the well known divergences in computing elements of S matrix. The ensuing Renormalization procedure began to cast the first doubts on the validity of the picture of particles as elementary pointlike objects. And, what is more important, made clear that the actual values of masses and electric charges of the elementary particles could not be derived theoretically from the first principles of some QFT-based model, but should be obtained only through experiments. In turn, this prompted a number of theorists to adopt a "phenomenological" attitude, by devoting themselves to building only specific models accounting for the data obtained from scattering experiments performed in large accelerators. Paradoxically people made use of a theory, like QFT, in which the concept of particle was not so clear in order to explain phenomena, like traces in bubble chambers, considered as trivially evidencing the particle-like nature of matter.

Such a situation gave rise to two different programs, each trying to remedy the conceptual difficulties of the original formulation of QFT: the *Effective Field Theories* (EFT) and the *Algebraic Quantum Field Theory* (AQFT). Within the former the Lagrangians and the coupling constants were viewed as dependent on the energy scale of the phenomena under consideration. On one side, this entails that each theory has a limited validity, associated to a specific energy range. On the other side, this allows to understand the divergences as resulting from the influence of higher energy processes (typically of microscopic nature) on lower energy ones (which can be considered as more macroscopic). It is important to remark that this view opens the way to a new kind of theories, in which physical constants, such as electron charge, Planck's constant, and like, are no more "sacred" quantities, but rather effective measures of the amount of interactions existing between different levels of observation. In more recent times, to quote an example, a number of researchers introduced an "effective Planck constant" (see, for instance, Artuso and Rusconi 2001; Averbukh *et al.* 2002). Such a concept has been very useful to set a connection between noisy PDE, such as noisy Burgers equation, and QFT (see Fogedby 1998; Fogedby and Brandenburg 2002). We can thus say that EFT approach is endowed with a number of remarkable potentialities, still largely unexplored (for conceptual analyses of the role played by EFT see Cao and Schweber, 1993; Hartmann, 2001; Castellani, 2002).

On the contrary, the program of AQFT, based on an abstract mathematical framework, takes into consideration, rather than fields, algebras of local observables, by supposing that all physical information of QFT is contained in the mapping $O \rightarrow A(O)$ from finite, open and bounded Minkowski spacetime regions $O$ to algebras $A(O)$ of local observables defined in $O$. This kind of approach originated from the work of Haag (see, among the earlier papers, Haag and Kastler, 1964) and, despite its uselessness in performing practical computations, appeared, already from its first introduction, as a very powerful tool to carry out deep conceptual analyses of the main foundational problems of QFT. Among the problems dealt with through the methods of AQFT (being impossible to quote the huge number of relevant references we will limit ourselves to mention some general introductions such as Horuzhy, 1990; Haag, 1996; Halvorson, 2006) we can quote the existence of unitarily inequivalent representations (granted by a celebrated theorem of Haag; see, e.g., Haag, 1961), the role of vacuum, the requirements of locality and causality as well as the relationships between QFT and Special and General Relativity Theories, and the localization of particles.

In particular, as regards the concept of particle and its localizability in QFT, the approach based on AQFT gave rise to a number of theorems which evidenced how QFT *cannot* be a theory of localizable particles, at least if it must avoid any contradiction with Special Relativity Theory. The first of these theorems is the so-called *Reeh-Schlieder theorem* (Reeh and Schlieder, 1961). It asserts that, in the case of a quantum system defined within a bounded open region $O$ of Minkowski space-time and associated with a suitable Hilbert space state $H$ as well as with a suitable Von Neumann algebra $A(O)$ of local observables (operators) defined in $O$, acting on system's vacuum state through elements of $A(O)$ one can approximate with a whatever precision any state of $H$, even if different from vacuum in some space-like separated region $O'$. This is equivalent to state that local measurements cannot allow any distinction between the vacuum state and, say, a $N$-particle state. The theorem can be interpreted in many ways (see, for instance, Redhead, 1995; Fleming, 2000; Clifton and Halvorson, 2001a; Halvorson, 2001), but undoubtedly it points to the fact that long-range correlations characterizing the vacuum state in quantum theories make impossible any interpretation of QFT as describing sets of interacting, pointlike or smoothly localized, particles.

In more recent times two further no-go theorems seemed to destroy any hope of describing localized particles within the framework of QFT. The first of them is the *Malament theorem* (Malament, 1996), holding for the general case of an affine space-time (hence not necessarily Minkowskian). The theorem makes use of the notion of *localization system*, based on a mapping $\Delta \mapsto E_\Delta$ from a bounded "spatial" region $\Delta$ to the proposition $E_\Delta$ asserting that a particle is localized in $\Delta$ with unit probability. Besides, it is supposed that the localization system satisfies the following four conditions:

1. *Localizability* ($E_\Delta E_{\Delta'} = 0$ if $\Delta$ and $\Delta'$ are disjoint spatial regions)
2. *Translation covariance* ($U(a)E_\Delta U(a)^* = E_{\Delta+a}$ for any $\Delta$ and any translation $a$)
3. *Energy bounded below*
4. *Microcausality* (if $\Delta$ and $\Delta'$ are disjoint spatial regions whose reciprocal distance is not zero, then for any timelike translation $a$ there is an $\varepsilon > 0$ such that $[E_\Delta, E_{\Delta'+ta}] = 0$ if $0 \leq t < \varepsilon$).

Then the Malament theorem asserts that, if the conditions 1-4 hold, $E_\Delta = 0$ for all $\Delta$. In other words, it is impossible to detect the particle in any spatial region. No localization is possible!

The second no-go theorem quoted above is the *Hegerfeldt theorem* (Hegerfeldt, 1998a; 1998b). Even in this case one deals with a localization system, but this time even the existence of a unitary time evolution operator $U_t$ is taken into account. The theorem asserts that if the localization system satisfies the following four conditions:

1a. *Monotonicity* (if a particle is localized in every one of a family of regions "approaching" $\Delta$, then it is localized in $\Delta$)
2a. *Time translation covariance* ($U_t E_\Delta U_{-t} = E_{\Delta+t}$ for any $\Delta$ and any $t$)
3a. *Energy bounded below*
4a. *No instantaneous wavepacket spreading,.*

then $U_t E_\Delta U_{-t} = E_\Delta$ for all $\Delta$ and all $t$. In other terms, if a particle is localized, no dynamics is possible (except the trivial one). Moreover, by adding the further condition of absence of an absolute velocity, it can be shown that the theorem entails $E_\Delta = 0$ for all $\Delta$ (see Halvorson and Clifton, 2002, Lemma 2 of Appendix A), like in Malament theorem.

The last class of results obtained within the framework of AQFT is due to the work of Clifton and Halvorson (see, for instance, Clifton and Halvorson, 2001b). These authors, exploiting the existence of unitarily inequivalent representations in QFT, were able to show that the same model can allow different inequivalent quantizations, each one corresponding to a different particle concept. In doing so, they resorted to the example given by Unruh effect (see Unruh, 1976; Unruh and Wald, 1984), consisting in the fact that a uniformly accelerated observer, lying within an empty Minkowski Universe, will detect a thermal bath of particles, named Rindler quanta. This effect has an intuitive explanation, stemming from the observation that a uniform acceleration entails the existence of a macroscopic force field (for instance gravitational), in turn giving rise to a curvature of spacetime. The latter, therefore, cannot be longer Minkowskian and the presence of curvature can originate an interaction between the different normal modes of the fields eventually present within spacetime, even if they are in a vacuum state (see, e.g., Wald, 1994; Arageorgis *et al.*, 2003). Such an interaction can result in a production of particles, which could be detected by an observer lying in a suitable reference frame (for instance uniformly accelerated). Of course, according to the principles of General Relativity Theory, the occurrence of this effect is strongly dependent on the kind of reference frame adopted and can take place or disappear as a function of allowed coordinate transformations, which could produce different inequivalent representations of the same QFT-based model. When dealing with QFT within curved spacetimes, therefore, even the concept itself of particle is devoid of any objective content.

To summarize, the possibility of a particle interpretation of QFT seems to have been ruled out. What is worrying is that almost all theorists continue to use QFT to describe particle behaviors, while probably their conclusions could not be grounded on a sound basis. On the other hand, could we renounce to electrons, protons or Quantum Electrodynamics? What to do in such a situation? In this regard we can identify three possible strategies for coming out of this impasse:

*a*) reformulate QFT only as a *field theory*, avoiding the concept of particle; of course such a strategy presupposes, as a counterpart, a classical theory of fields in which the concept of particle should, eventually, be introduced only to characterize the regions in which field strength is particularly high; the implementation of such a program, which we could label as the realization of the *Maxwellian dream*, should, of course, avoid any kind of divergences without, at the same time, introducing other entities extraneous to field themselves: in a sense we should still work within a *closed* world (eventually allowing suitable boundary conditions at infinity);

*b*) modify the mechanism actually used in QFT to reach a particle interpretation, based on normal mode decomposition, in such a way as to account for the nonlinear descriptions of interacting fields; the concept of particle should in some way be associated with "generalized" normal modes avoiding the problems arising when we try to extend the free field formalism to the interaction case; while keeping Lorentz invariance would be desirable, this strategy could allow, in particular contexts, also different kinds of invariance; after all, nobody would be surprised if, very close to a particle, the spacetime would deviate from Minkowskian form;

*c*) reformulate QFT as a theory of open systems interacting with a suitable environment; the new theory should, of course, renounce to concepts such as equilibrium states, ground states, exact invariance; the framework of EFT could, undoubtedly, fit in very well with this program; it is to be expected that the concept of particle, if any, should emerge from a concurrence of different factors, some of which typically contextual; in any case it should be endowed with a variability unknown to actual models of QFT.

The remainder of this paper will be devoted to a short discussion of the feasibility of each one of these programs (not mutually excluding, however). The aim will not be the one of proposing the "correct" solution to the problems mentioned above (a solution which, in principle, could not exist), but rather of listing the obstacles , already found or expected to be found, which prevent from implementing in an easy and satisfactory way the ideas underlying the programs themselves. Only in this way it is possible to understand whether the actual form of QFT represents only a step towards the building of the more ambitious theory human beings can conceive or is the disastrous conclusion of a crazy intellectual adventure.

**2. The Maxwellian dream**
To start, let us shortly recall the main requirements to be satisfied in order to concretely implement the Maxwellian dream:
*r.1*) we should find a classical field theory such that the solutions of the associated field equations were always functions free from singularities; hereafter we will denote such a singularity-free field theory by SFFT;
*r.2*) SFFT should not be in contrast with experimental data; thus, it cannot differ very much from the field theories usually adopted, like, for instance, the Maxwell electromagnetic field theory;
*r.3*) among the solutions of SFFT in absence of sources there should be some "bump-like" ones, of course associated with finite energies;
*r.4*) the bump-like solutions should behave, more or less, like localized (but not pointlike) particles; the latter should be associated to suitable "charges", in turn expressed in terms of field strengths; moreover, we should allow dynamic solutions describing moving bump-like solutions, behaving in a soliton-like manner;
*r.5*) SFFT should be readily quantizable; besides, its quantum version should be free from infinities, thus avoiding any need of renormalization.
At first sight it would seem that a SFFT satisfying at least some of the previous requirements really exists. We speak here of Born-Infeld electromagnetic field theory (Born, 1933; 1934; Born and Infeld, 1933; 1934). As well known, the latter is a non-linear generalization of the Maxwell electromagnetic field theory, whose Lagrangian was originally written under the form:

$$(1) \quad L = b^2 \left[ 1 - \sqrt{1 + (F_{\mu\nu} F^{\mu\nu} / 2b^2)} \right]$$

where $F_{\mu\nu}$ denotes the usual electromagnetic field tensor, $b$ is a positive parameter, and the tensor indices are raised or lowered in conformity with a Minkowski metric $\eta_{\mu\nu} = diag(+1,-1,-1,-1)$. Born and Infeld, relying on invariance considerations, introduced also a further generalization of (1), which can be written as:

$$(2) \quad L' = b^2 \left\{ 1 - \sqrt{-\det[\eta_{\mu\nu} + (F_{\mu\nu} / b)]} \right\}$$

The Lagrangian (2) contains further non-linear terms with respect to (1). Namely its explicit form is:

$$(3) \quad L' = b^2 \left\{ 1 - \sqrt{1 + (F_{\mu\nu} F^{\mu\nu} / 2b^2) - [(F_{\mu\nu} * F^{\mu\nu})^2 / 16b^4]} \right\}$$

where $*F^{\mu\nu} = \frac{1}{2} \varepsilon^{\mu\nu\alpha\beta} F_{\alpha\beta}$. So far there is no general consensus about what Lagrangian, (1) or (2), should be more convenient. In any case, for most static problems the choice of (1) or (2) makes no difference. It is easy to see that, when $b \to +\infty$, the Lagrangian (1) tends to the usual Maxwell Lagrangian:

$$(4) \quad L_M = -\frac{1}{4} F_{\mu\nu} F^{\mu\nu}$$

Thus, the requirement *r.2)* would seem satisfied. As regards the requirement *r.1)* there is some indication that even the latter could be satisfied, provided, however, that we allow some "breaking" of the original Maxwellian dream. Namely, if we search for the solution of field equations deriving from (1) in the case of the spherically symmetric electric field produced (alas!) by a pointlike charge, it can be found that the scalar potential is given by:

$$\varphi = \frac{e}{r_0} \int_{r/r_0}^{\infty} d\xi \frac{1}{\sqrt{1+\xi^4}} \tag{5}$$

where $e$ denotes the value of the charge, $r$ is a radial coordinate (the charge is located in $r=0$), and $r_0$ is given by:

$$r_0 = \sqrt{\frac{|e|}{b}} \tag{6}$$

It can be easily seen that, when $r$ is far greater than $r_0$, $\varphi$ behaves like $e/r$, like in Maxwellian case, while, when $r \to 0$, $\varphi$ tends to the limiting value $1.8541(e/r_0)$, which is finite. Thus, contrarily to what occurs in Maxwell's theory, there is no divergence in correspondence to the charge. This can be seen also in the formula, derived from (5), which gives the electric field:

$$E = \frac{e}{r} \frac{\vec{r}}{\sqrt{r^4 + r_0^4}} \tag{7}$$

When $r=0$ the value of $E$ is finite and given by $b$. As a consequence the energy is always finite and the infinities troubling Maxwellian theory seem to have disappeared.

But, can we consider Born-Infeld theory as a good candidate for a SFFT? Unfortunately not. Namely it can be rigorously proved (see Yang, 2000) that this theory allow only one static sourceless solution, corresponding to zero electric and magnetic field. Therefore, no bump-like field configurations is possible in absence of pointlike sources. What to say about non-static solutions? It is known that this theory allow the propagation of non-linear electromagnetic waves (see, e.g., Bialynicki-Birula, 1983). These waves are characterized by absence of birefringence, that is the propagation occurs along a single light-cone, as well as by absence of shock waves. Approximate computations of the waveforms as well as of the dispersion relationships have already been performed in correspondence to the choice of the Lagrangian (2), a circumstance which opens the way to possible experimental test of the validity of Born-Infeld theory (see, for instance, Denisov, 2000; Ferraro, 2007). As regards, instead, the existence of solitons we still lack definite results. When the Born-Infeld field is coupled to other fields we obtain complicated nonlinear systems which sometimes allow solitonic solutions. However, the presence of solitons is always due to the other field and not to the Born-Infeld one. A typical case is the one of a Klein-Gordon charged scalar field which, when coupled to a self-produced electromagnetic field, allows solitonic solutions even when the latter is described by the usual Maxwell equations (see Long and Stuart, 2009). Thus, one would not be surprised if the same phenomenon should occur even if the electromagnetic field were described by Born-Infeld theory. We add here a further, even if obvious, remark: Born-Infeld equations probably could easily allow solitonic solutions (good candidates for particles) if we supposed that the vacuum were consisting in a highly nonlinear medium. This effect already occurs for the Maxwell electromagnetic equations which, while being linear, allow solitonic solutions in the case of propagation within nonlinear media with special properties (see, e.g., Snyder and Mitchell, 1998). Without speculating about the introduction of new strange kinds of aether, the previous remark has been made to stress the fact that probably Born-Infeld theory will never be able to generate in autonomous way localized soliton-like solutions in absence of sources without the help of another field.

The above considerations evidence that the Born-Infeld theory is unable to satisfy the requirements *r.3*) and *r.4*). The Maxwellian dream, therefore, cannot be implemented in this way. To ending our discussion, however, we will spend some words about the quantization of Born-Infeld theory. As suspected, it is very far from being easy and very few authors dealt with problems (see Hotta *et al.*, 2004; Kogut and Sinclair, 2006). In all cases, owing to the impossibility of using traditional methods, it needed to resort to heavy numerical simulations, whose interpretation is always not so reliable. However they gave a mild indication that a quantum Born-Infeld theory resembles in some way to a Quantum Electrodynamics with "heavy" photons. Unfortunately here the massive photons have a mass which varies as a function of the field itself. This can be easily seen when looking at the equations which describe the wave propagation. These latter, if we use the Lagrangian (1), have the form:

$$(8) \qquad \partial_\mu F^{\mu\nu} = (\partial_\mu \ln \chi) F^{\mu\nu} \qquad \chi = \left[ \sqrt{1 + (F_{\mu\nu} F^{\mu\nu}/2b^2)} \right]$$

In the case of a static field given by (7), straightforward computations show that, by supposing that the value of *b* be high enough, the mass $m_{BI}$ of the Born-Infeld "photon", when *r* is far greater than $r_0$, behaves approximately according to a law of the form:

$$(9) \qquad m_{BI} = (\partial_\mu \ln \chi) = \frac{2e^2}{b^2} \frac{1}{r^5}$$

To summarize, even the requirement *r.5*) cannot satisfied and we are forced to conclude that the Maxwellian dream has no hope of being realized through the introduction of Born-Infeld theory. We could, of course, object that Born-Infeld theory is not the ultimate one: other theories allow solitonic solutions and localized objects of any sort. This is a very active field of research (while it is impossible even to limit ourselves to quote only the most relevant references within an enormous amount of literature, we will be forced to mention only few review papers, such as Belova and Kudryavtsev, 1997; Maccari, 2006; Manton, 2008) and a number of people hold that solitons could give a concrete alternative to formulate a concept of particle better than the one stemming from the usual quantization methods. The latter idea, however, appears very difficult to support, for three main reasons:

i) almost all field equations allowing solitonic solutions have little or no relationship with the ones used in QFT to describe the fundamental force fields; equations like Sine-Gordon, Nonlinear Schrödinger, Kadomtsev-Petriashvili, while describing the phenomenological behaviors of a number of physical systems, still constitute *ad hoc* equations, in which often the solitonic behaviors arise as a byproduct of special boundary conditions or of particular choices of parameter values; while it is undeniable that some of these equations can be obtained as particular approximations of more general equations describing fundamental interactions, we cannot forget that, in introducing these approximations, we just lost the generality required for a QFT-based model;

ii) some of solitonic solutions are unstable; as regards the majority of the ones so far found (often through numerical methods) there is no proof neither of their stability, nor of their instability; this applies chiefly to solitons in (3+1)-dimensional spacetime, the most suited ones to describe physically realistic particle models; moreover, numerical experiments and theoretical considerations evidenced that, when two solitons collide, often they lose their individuality, a circumstance which cast serious doubts on the usefulness of solitons as models of particles;

iii) in general, quantization of solitons is difficult, even if a number of approximate methods are in use; this justifies the name "quantum solitons", even if in most cases it refers to normal mode expansions of linear approximations of excitations.

The considerations made within this paragraph, therefore, let us understand that the Maxwellian dream is very far from its realization, both in the case we believe in Born-Infeld theory and in the case we believe in solitons. This lead us to explore alternative ways for defining the concept of particle in QFT.

**3. The non-linear generalization of QFT**
As told in every textbook, usually in QFT the concept of particle is introduced via the construction of Fock space representation of equal-time canonical commutation (CCR) relations for a free field. In the case of a bosonic field this is based on the introduction of suitable creation and annihilation operators $a^\dagger(k,t)$ and $a(k,t)$ obeying the CCR:

$$(10) \quad [a(k,t), a(k',t)] = [a^\dagger(k,t), a^\dagger(k',t)] = 0 \quad, \quad [a(k,t), a^\dagger(k',t)] = \delta^3(k-k')$$

so that, in the case of a scalar quantum field $\phi(x,t)$, we can represent it under the form:

$$(11) \quad \phi(x,t) = \int \frac{d^3k}{(2\pi)^{3/2}\sqrt{2\omega_k}} \left[ a^\dagger(k,t) e^{ikx} + a(k,t) e^{-ikx} \right]$$

where:

$$(12) \quad \omega_k^2 = k^2 + m^2$$

It is to be recalled that (11) is nothing but a Fourier decomposition into positive and negative frequency modes. In this regard we remark that Fourier decomposition and the whole Fourier analysis is a typically linear tool, useful for linear systems but not so suited to deal with non-linear systems, such as interacting fields. Thus the identification of these modes with "quanta" whose mass is *m* makes sense only within a linear context, such as the one of free fields. Now, to complete this identification, we need to introduce a no-particle state $|0\rangle$, usually identified with the *ground state* or *vacuum state* (supposed unique), such that, for all $k$, we have:

$$(13) \quad a(k,t)|0\rangle = 0$$

This allows to introduce *n*-particle states by acting *n* times on the vacuum through the creation operator $a^\dagger(k,t)$. While neglecting here all details related to the normalization of these states and to the smearing of the operators, we will limit ourselves to mention that, by taking the direct sum of the *n*-fold symmetric tensor products of one-particle Hilbert spaces, we will obtain the Fock space of the free field taken into consideration. The Fock space thus defined is freely used by the majority of theorists applying QFT to describe interacting particles. However, the above observations induce to suspect that this use is incorrect and in the following we will present some arguments which corroborate this suspicion.

A first argument (here we will partly follow the line of reasoning adopted in Fraser, 2008) comes from the Haag theorem, asserting the existence of unitarily inequivalent representations of CCR in QFT (see, in this regard, also Earman and Fraser, 2006). In particular, the representations associated with interacting fields are not equivalent to the ones associated with free fields, otherwise we should have unitary transformations producing the disappearance of interaction, letting us go from the description of interaction to the description in absence of interaction without changing the physical content of the theory. As a consequence the description of quanta obtained through (10)-(13) in the case of a free field can no longer hold in the case of interacting fields. In other terms, the particle description associated with the free field must be *physically different* from the one (if any) holding in presence of interaction.

A second argument is related to the fact that, in presence of interactions, the uniqueness of vacuum cannot be longer granted. Let us think, for instance, of spontaneous symmetry breaking when we are in presence of many inequivalent vacua. As well known, in this case it is very difficult to describe what occurs close to the critical point of the transition, where the concept itself of particle has only an heuristic value, and one needs to resort to approximate methods, which allow to perform concrete computations (see, among the others, Umezawa, 1993; Vitiello, 2005; Del Giudice and Vitiello, 2006; Pessa, 2008). The concept of particle can be reintroduced, in such situations, only if representations of CCR are related to *asymptotic states*, occurring when the interactions have been turned off.

A third argument concerns the possibility of introducing a decomposition of the field operator different from the Fourier one, but more suited to the needs of nonlinear descriptions of interacting fields. Unfortunately, so far such a decomposition has been not found. Of course, there exist many different methods, widely used in data and signal analysis (e.g. wavelet transforms) but none of them can warrant Lorentz invariance nor the possibility of obtaining relationships such as (13). In absence of these characteristics, it is evident that the traditional particle interpretation is untenable.

As a consequence of these arguments it seems that a slight modification of the traditional formalism of QFT, keeping unchanged the main structure of this theory, cannot improve the situation, at least as regards the introduction of an acceptable particle concept. This, in ultimate analysis, stems form the intrinsic nonlocality associated with both QM and QFT which frustrates any effort to define a localization operator endowed with acceptable properties. Such a difficulty persists even if we introduce, by hand, suitable pointlike sources linearly interacting with a quantum field, as the initially localized quantum states inevitably spread as a consequence of field dynamical evolution (see, e.g., Buscemi and Compagno, 2006; as regards the quantum delocalization of electric charge see Buchholz *et al.*, 2001).

Such a situation seems to call for a radical revision of QFT, as well as of the intuitive conception of particle, so far viewed as a sort of strongly localized object, almost pointlike, endowed with an inner invariance as regards its main characteristics. Such a revision, of course, needs to interact with other theoretical constructs, such as General Relativity Theory, whose relationships with QFT have been, so far, rather difficult. And it is not so strange that this enterprise has been set up by string theorists when attempting to build a general theory of quantum gravity. But in this context the actual situation seems so complex as to hinder any theoretical fallout, at least as regards the concept of particle in QFT. Some inspiration, however, can come from another domain, whose relationships with QFT have been very fruitful: the one of condensed matter. Here one of the most interesting concepts is the one of quasi-particle, consisting in a collective excitation emerging from the local interactions between the elementary constituents of a complex system (like, for instance, phonons in a crystal). Quasi-particles share with traditional particles many features, except localization. However, in presence of suitable contexts, they can give origin to localized entities under the form of travelling solitons (like in the Davydov effect; see, e.g., Davydov, 1979; 1982; Scott, 1992; 2003; Förner, 1997; Brizhik *et al.*, 2004; a discussion of this effect within QFT is contained in Del Giudice *et al.*, 1985). This seems to point towards a more general conception of particle as a sort of "emergent effect", whose practical description, in terms of a suitable EFT, could also exhibit some of the aspects commonly associated with the traditional views. The need for such a framework has been emphasized, for instance, by Wallace (Wallace, 2001) and Zeh (Zeh, 2003). On the other hand, we remind that within QFT already exist techniques to deal with description of particles as "collective effects" (see, e.g., Novozhilov and Novozhilov, 2001). However, the practical implementation of this idea does not appear as so simple. In the following we shortly describe the principles underlying a possible attempt to map the QFT formalism on the simplest world of discretized lattice models, while avoiding the complexity and the computational costs associated with Quantum Monte Carlo simulations in lattice gauge theory. These kinds of mapping can be useful to make available to the widest possible audience tools which allow to concretely follow the dynamical evolution in time of field quantities, so as to understand what practical meaning and what limits would have the notion of "particle" in real contexts.

## 4. A discretization method for dynamical field equations

Discretization methods are widely used in numerical analysis, as well as in many models of complex systems, including artificial neural networks, agent models, cellular automata, social networks, and so on. However, they work even in contexts such as quantum gravity (see, e.g., Zizzi, 1999; 2008) owing to the unavoidable spacetime quantization at the level of Planck's length. As regards these methods, there is a great variety of possible choices. Here we will illustrate a possible path towards the discretization of the dynamical equations driving the evolution of a quantum field, based on the following steps:

*s.1*) let us begin by discretizing the spatial variables, so as to express spatial derivatives through finite differences; this implies that the space has to be considered as a lattice of sites (possibly regularly spaced); in this way we deal only with a set of ordinary differential equations for the field operators;

*s.2*) let us look at the classical version of this set of differential equations, trying to find a set of classical solutions of them; let us discretize this set (for instance by resorting to the parameter values appearing within them) and choose a finite subset of it; moreover, it is advisable to exploit the symmetries and the invariances of the classical equations so as to subdivide the set of solutions into two subsets, in such a way as to have the possibility of mapping each solution belonging to one of these subsets into the corresponding solution belonging to the other subset through a simple symmetry transformation; to make a simple example, in the case of field equations in which the only differential operator is given by the Dalembertian, if $u(x,t)$ is a solution of field equations, even $u(x,-t)$ will be a solution, so the two subsets of solutions can be transformed one into the other by simply reversing the sign of time;

*s.3*) let us express each field operator through a linear (and finite) combination of the classical solutions whose weights are operators, dependent on time and on the chosen site, acting as creation and annihilation operators of the associated solutions in the site under consideration; these operators should be considered as Heisenberg operators, which can vary with time under the action of field Hamiltonian; they help to measure the probability that at a given time and in a given site, a projective measure performed around the site (we could say *on the site*, but from the point of quantum theory this is an excessive idealization, as we could not know a location with infinite precision) can give as outcome the value of the associated classical function in that site;

*s.4*) let us substitute this development into the field equations such as to obtain dynamical equations ruling the time evolution of the creation and annihilation operators;

*s.5*) let us use these equations to derive the associated equations ruling the time evolution of the probabilities of the different classical solutions for each site;

*s.6*) let us solve numerically the latter equations and observe the time evolution of the system; we are looking at a (somewhat rough) simulation of the dynamics of a quantum field; if there is something which could be interpreted as a particle, we could decide why this identification is acceptable or why not; in this way we could perhaps learn about concrete particles much more than looking at abstract models.

In order to illustrate this procedure through a toy example, we will resort to the model of a self-interacting (1+1)-dimensional Klein-Gordon scalar field described by:

(14) $$\partial_{tt}\varphi - \partial_{xx}\varphi + m^2\varphi = -4\lambda\varphi^3$$

As well known, the classical counterpart of (14) allows, among the others, soliton solutions consisting in kinks moving at a velocity $V$ and represented by functions of the form:

(15) $$\varphi_{x_0,V}(x,t) = \frac{m}{\sqrt{4\lambda}} \tanh\left[\frac{m(x-x_0-Vt)}{\sqrt{2(1-V^2)}}\right]$$

These solutions are parametrized by the values of $x_0$ and $V$. Moreover, it is immediate to see that, if $\varphi_{x_0,V}(x,t)$ is a solution, even $\varphi_{x_0,V}(x,-t)$ is a solution of the same equations. Let us now spatially discretize the field equation (14) on a 1-dimensional lattice whose discretization step is $h$ so as to obtain the following system of ordinary differential equations:

(16) $$\frac{d^2\varphi_i}{dt^2} = \frac{1}{h^2}(\varphi_{i+1}+\varphi_{i-1}) + \mu\varphi_i - 4\lambda\varphi_i^3 \quad , \quad \mu = m^2 + \frac{2}{h^2}$$

Here $i$ denotes the site index, while the operators $\varphi_i$ are to be considered as time-varying operators, in the Heisenberg picture, acting on the $i$-th site. Let us now introduce the development:

(17) $$\varphi_i = \sum_j \chi_j(i,t)a^\dagger_j(i,t) + \chi_j(i,-t)a_j(i,t)$$

where the symbols $\chi_j(i,t)$ denote c-number functions of the form (15) and, for saving space, the parameters have been synthetized by a single index $j$. The operators $a^\dagger_j(i,t)$ and $a_j(i,t)$ are respectively creation and annihilation operators for the functions $\chi_j(i,t)$, acting on the $i$-th site. More precisely, $a^\dagger_j(i,t)$ allows $\chi_j(i,t)$ as its eigenfunction with eigenvalue $\alpha_j(i,t)$ and, when acting on $\chi_j(i,-t)$, transforms it into $\chi_j(i,t)$. On the contrary $a_j(i,t)$ allows $\chi_j(i,-t)$ as its eigenfunction with eigenvalue $\beta_j(i,t)$ and, when acting on $\chi_j(i,t)$, transforms it into $\chi_j(i,-t)$. We stress here the deep conceptual difference between the development (17) and the traditional one given by (14). The latter describes a decomposition in terms of harmonic oscillators while (17) makes use of basic functions naturally suited to the problem at hand, so that the nonlinearity itself is embedded from the starting within the formalism.

Trivial considerations show that these operators fulfil the commutation relations:

(18) $$[a_j(i,t), a_j(i,t)] = [a^\dagger_j(i,t), a^\dagger_j(i,t)] = 0 \quad , \quad [a_j(i,t), a^\dagger_j(i,t)] = \beta_j(i,t)\chi_j(i,-t) - \alpha_j(i,t)\chi_j(i,t)$$

If we substitute the development (17) into (16) straightforward computations allow to obtain the following equations ruling the time evolutions of the operators $a^\dagger_j(i,t)$ and $a_j(i,t)$:

(19.a) $$\frac{d^2 a^\dagger_j(i,t)}{dt^2} = \frac{1}{h^2}\left[\chi_j(i+1,t)a^\dagger_j(i+1,t) + \chi_j(i-1,t)a^\dagger_j(i-1,t)\right] - Q_j(i,t)a^\dagger_j(i,t)$$

(19.b) $$\frac{d^2 a_j(i,t)}{dt^2} = \frac{1}{h^2}\left[\chi_j(i+1,-t)a_j(i+1,t) + \chi_j(i-1,-t)a_j(i-1,t)\right] - R_j(i,t)a_j(i,t)$$

where:

(20.a) $$Q_j(i,t) = \frac{1}{h^2}\left[\chi_j(i+1,t) + \chi_j(i-1,t)\right] + 4\lambda\left[2\chi_j(i,t)\chi_j(i,-t) + \chi_j^2(i,-t)\right]$$

(20.b) $$R_j(i,t) = \frac{1}{h^2}\left[\chi_j(i+1,-t) + \chi_j(i-1,-t)\right] + 4\lambda\left[2\chi_j(i,t)\chi_j(i,-t) + \chi_j^2(i,t)\right]$$

From the definitions themselves of the operators it follows that their eigenvalues are directly proportional to the probabilities of occurrence of the behaviors described by the c-number functions (that is the eigenfunctions) to which they are associated. If, now, we assume that these eigenvalues are normalized, that is they sum up to 1 on all possible values of $j$, then, by applying both members of the (19.a) and (19.b) to the states described by the respective eigenfunctions of these operators, and taking the expectation values of the operators themselves, it is possible to derive from these equations, previous suitable normalizations, that the probabilities of occurrence of the states described by the functions $\chi_j(i,t)$ and $\chi_j(i,-t)$, denoted respectively by $p_j(i,t)$ and $q_j(i,t)$, satisfy classical differential equations formally identical to the (19) holding for $a^\dagger_j(i,t)$ and $a_j(i,t)$. It is then possible to simulate on a computer the time evolution of this discretized version of a quantum field theory, provided that, at each time step, once updated for each site the values of the probabilities of occurrence of the states corresponding to the different values of $j$, we renormalize these values (by dividing by their sum) so as to keep invariant the fact that they must sum up to 1. The updating, of course, can be easily implemented through a routine for a numerical integration of the equations (19). In the following Figures 1.a and 1.b we show two different snapshots of a particular evolution on a lattice of 100 sites, with parameter values $m=1$, $\lambda = 0.25$, corresponding, respectively, to the steps $t=10$ and $t=500$. Nedless, to say, the initial state was chosen as given by a function of the form (15).

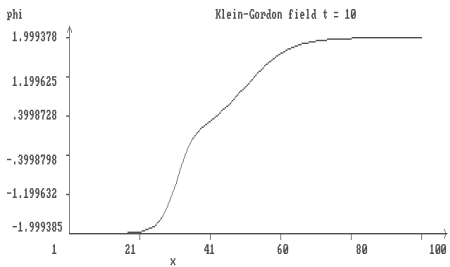   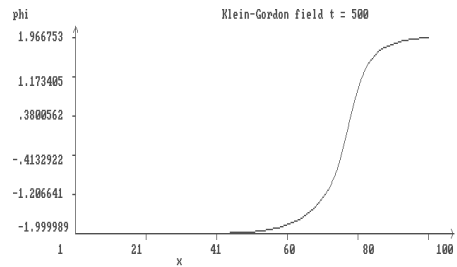

Figure 1.a                                   Figure 1.b
$\varphi_i$ vs $i$ when $t=10$           $\varphi_i$ vs $i$ when $t=500$

Of course, being our field endowed with a quantum nature, another snapshot, taken in the identical conditions, could give rise to very different trends. Namely, the time evolution being stochastic, it is based on drawing random numbers. Eventual conclusions, if any, could be reliable only after a careful statistical analysis of a large number of numerical simulations. Anyway it is possible to easily detect in both Figures the presence of the "particle" of our system, that is the moving kink. What, however, attracts our attention is that the kink, while keeping more or less unchanged its general form, undergoes small scale changes, like fluctuations, particularly evident in Figure 1.a. This raises a conceptual problem: how much fluctuations can we tolerate in order to consider a particle as an almost invariant entity, endowed with a specific identity? Could we identify, for instance, a particle with some kind of statistical construct related to empirical data? Is the concept of particle a fuzzy concept? Could we, up to a certain degree, deal with particles in the same way as the zoologists deal with animal species?

The answer to these questions is largely dependent on the goals and the cultural background of physicists and/or philosophers. It is to be supposed that condensed matter physicists, as well as experimental physicists, would be more inclined to accept a tolerant view about the concept of particle, considered mostly as an heuristic tool to build an EFT and to account for some sets of experimental data. On the other hand, physicists dealing with the theory of fundamental interactions, as well as philosophers, would be uncomfortable in presence of concepts not exactly defined, fearing the collapse of theories, like the Standard Model, which required a tremendous intellectual effort. The latter categories of researchers will never be satisfied with numerical simulations. They will always search for a strongly grounded theoretical apparatus, unassailable from every point of view. Unfortunately (or luckily) theoretical physics cannot offer such products.

**5. QFT as a theory of open systems**
Before ending our considerations we will briefly mention a different research program, trying to generalize QFT in such a way as to describe quantum fields interacting with a suitable external environment. At first sight this program seems incompatible with the general principles which guided from the starting the building of QFT. Namely, as this theory was dealing with fields, conceived as entities filling the whole spacetime without boundaries, it was viewed as a theory of *closed* systems. Therefore the attempt to use QFT to describe *open* systems seem to run into contradiction. Nevertheless, the need for such a generalization came from a number of different domains:

*d.1*) most macroscopic systems are influenced by noise; the latter acts like a sort of external source able, in some cases, to destroy quantum coherence and, in others, to give rise to new structures which would disappear in a noiseless situation; without taking into account a noisy environment QFT could be applied only to very high energy physics;

*d.2*) when a quantum field interacts with a classical field, such as, for instance, the gravitational field, it is possible to have different forms of backreaction, giving rise, among the others, to particle production; thus within a curved spacetime it is unavoidable to consider a quantum field as endowed with an (active) environment interacting with it;

*d.3*) phenomena such as spontaneous symmetry breaking or phase transitions could not be possible without the changes occurring in the values of suitable critical parameters, in turn triggered by the action of some kind of external environment;

*d.4*) the emergence of macroscopic quantum effects can, in principle, be controlled (eventually resorting to some form of quantum control) through suitable influences exerted by an external environment; of course, to describe such phenomena we need a very sophisticated version of QFT, as we must take into account the possibility of metastable states, multiple vacua, and so on.

To generalize QFT in order to deal with these problems we need to resort to a statistical approach focussed on correlation functions and on the study of fluctuations. Within this context (see, for instance, Calzetta and Hu, 2000; 2008) it is more convenient to rewrite the field dynamical equations under the so-called Schwinger-Dyson form, which lets us express the propagators in terms of the other correlation functions of the field. It allows to evidence a feature typical of interacting fields: the two-point correlators are dependent on higher order correlators. A simple example is given by the self-interacting Klein-Gordon field described by (14). In this case the Schwinger-Dyson equation assumes the form:

(21) $$(\partial_{xx} - \partial_{tt} - m^2)G_F(\xi,\xi') = -i\delta(t-t') + 4\lambda\langle T|\varphi^3(\xi)\varphi(\xi')|\rangle$$

where the symbols $\xi$, $\xi'$ denote spacetime points, $T$ is the usual chronological ordering operator, and $G_F(\xi,\xi')$ is the usual Feynman propagator defined by:

(22) $$G_F(\xi,\xi') = \langle T|\varphi(\xi)\varphi(\xi')|\rangle$$

As it is easy to see from (21), the dynamics of the two-point function $G_F(\xi,\xi')$ depends in turn on a fourth order correlator. And the dynamics of the latter will depend on highest order correlators, and so on. This recursion *ad infinitum* naturally calls for some truncation prescription in order to obtain the effective dynamics. But, as it easy to understand, such a truncation will inevitably leave out some contributions to dynamics, which we will be forced to consider as fluctuations and therefore as a sort of noise. Thus, the latter is an unavoidable ingredient of QFT.

This situation becomes worse in presence of an external source, chiefly if the latter includes a noisy contribution. Namely the latter can act on the correlators at every order, thus inducing a deep modification of the nature itself of quantum fluctuations. If, now, we consider particles as emergent from field fluctuations, it becomes evident that, in presence of an external environment, we will be forced to take into account the contribution to these fluctuations given by the environment itself, in turn dependent on its correlation functions. Thus, once accepted the idea of an open version of QFT, the concept of particle becomes a byproduct of the interactions between the fields and the environment. We could therefore say that a particle is nothing but an effective field description of a complicated dynamics coupling the environment and the fields, a description therefore embodying not only the description of fields, but even the one of environment.

Despite the attractiveness of this approach, its practical implementation is, however, very difficult, mainly for technical reasons stemming from the fact that the actual mathematical structure of QFT has been built to describe only free fields, generalizing the usual tools of classical and quantum mechanics. Another difficulty follows from the fact that so far we still lack acceptable and realistic descriptions of the possible environments, notwithstanding the evidence concerning the deep influence of environment structural features on the evolution of quantum systems (see, for instance, Zurek, 2003; Montina and Arecchi, 2008). This is witnessed by the fact that only in the last years the concept of dissipation has been included within QFT, mostly through a mechanism of *doubling* of degrees of freedom (for more details we refer to Celeghini *et al.*, 1992; Vitiello, 2001; Blasone *et al.*, 2001; 2005; 2006). But there is the hope that the combined effort of those contributing to introduce new theoretical and mathematical constructs and of those building numerical models, such as the one sketched in the previous section, will result in significant new acquisitions.

## 6. Conclusions

After this complicated trip in the endless field of theoretical physics, we still are in a state of uncertainty. The naïve concept of particle, adopted by most practitioners of QFT, evidences intrinsic contradictions and therefore should be abandoned. This in turn implies a deep reformulation of the whole apparatus of QFT. In this regard, however, all proposals so far made are plagued by serious shortcomings which, so far, prevents from the introduction of a new, and more firmly grounded, concept of particle. It seems, after all, that we do not need a rigorous definition of the latter. QFT can work and produce acceptable previsions even in absence of it. Nevertheless, from a practical point of view, we need to summarize a number of experimental facts and theoretical features by introducing the concept of particle which, undoubtedly, allows more economical descriptions and more easily understandable pictures of dynamical phenomenology. Within this context, we can be satisfied with a definition of particle as a construct having a citizenship within an effective field theory, more or less like quasi-particles. As such, this construct must necessarily be endowed with dynamical features, which were absent in the old models of pointlike particles. Of course, the technical ingredients needed to introduce the new "effective" definition of particle are still incomplete and lot of work is necessary before obtaining significant advances along this direction. While this situation is satisfactory for most physicists, we acknowledge that it could be embarrassing for those searching for the "fundamental particles". However, nobody prevents from thinking that, at very high energy, the "effective" description of particles will reduce to the one of (almost) pointlike particles. And most actual efforts of theoretical as well as experimental physicists try just to prove the validity of this hypothesis. The ones which will remain unsatisfied for this state of affairs are the philosophers (or at least some of them). Namely the solution we have sketched above entails the disappearance of the *haecceitas* of particles, which are reduced to mere auxiliary constructs, useful for practical purposes, but in turn making reference to deeper constructs. For these philosophers the problem now becomes: what are these constructs? Do they coincide with fields? In this regard there are already some indications about a possible negative answer to this question (Teller, 1990; 1995). Perhaps, as suggested by Cao (see, Cao, 1997; 1999), the best ontological basis for QFT is given by its structure itself (inextricably connected with the processes it describes) rather than by specific entities (particles or fields).